\documentstyle[twoside,fleqn,npb,epsfig]{article}
\newcommand{\AmS}{{\protect\the\textfont2
  A\kern-.1667em\lower.5ex\hbox{M}\kern-.125emS}}

\title{The AGL Equation from a  Dipole Picture}

\author{ M. B. Gay  Ducati and V. P. Gon\c{c}alves  \address{Instituto de F\'{\i}sica, Univ. Federal do Rio Grande do Sul \\
 Caixa Postal 15051, 91501-970 Porto Alegre, RS, BRAZIL}
}

\begin{document}

\begin{abstract}

The AGL equation includes all multiple pomeron exchanges in the double logarithmic approximation (DLA) limit, leading to an unitarized gluon distribution in the small $x$ regime. This equation was originally obtained using the Glauber-Mueller approach. We demonstrate in this contribution that the AGL equation can also  be obtained  from the   dipole picture.  Our conclusion is that the AGL equation is a good candidate for an unitarized evolution equation at small $x$ in the DLA limit.

\end{abstract}

\maketitle

\section{INTRODUCTION}

The behavior of the cross sections in the high energy limit ($s \rightarrow \infty$) and fixed momentum transfer is expected to be described by the BFKL equation. The simplest process where  this equation applies  is the high energy scattering between two heavy quark-antiquark states, {\it i.e.} the onium-onium  scattering.
This process was studied in the dipole picture \cite{mueller},  where the 
heavy quark-antiquark pair and the 
soft gluons in the limit of large number of colors  $N_c$  are viewed as a collection of color 
dipoles. 
One of the main characteristics of the BFKL equation is that it predicts 
very high density of partons  in the small $x$ region.
 Therefore a new dynamical effect  associated with the unitarity corrections is expected to stop further growth of the parton densities. The understanding of the unitarity corrections has been a challenge of perturbative QCD (PQCD). 

Recently, an eikonal approach to  the  
unitarity corrections was proposed in the literature \cite{ayala1}. 
 The starting point of this paper is the proof of the Glauber formula in QCD \cite{muegla}, which considers only the interaction  of the fastest partons  with the target.
In \cite{ayala1}, a generalized equation which takes into account the interaction 
of all partons in a parton cascade with the target in the DLA limit  was proposed by Ayala, Gay Ducati, and Levin (AGL). 
The main properties of the  generalized equation  are that (i) the iterations of this 
equation coincide with the iteration of the Glauber-Mueller formula; 
(ii) its solution matches  the solution of the DGLAP evolution equation 
in the DLA limit of PQCD; (iii) it 
has the GLR equation as a  limit, and (iv)  contains the 
Glauber-Mueller formula. Therefore, the AGL equation is valid in a 
large kinematic region.  

The AGL equation resums all multiple pomeron exchanges in the DLA limit. Its  assintotic solution is given by $xG \propto Q^2\,R^2\,ln\,(1/x)$, where $R$ is the size of the target,  {\it i.e.} differently from the GLR equation, it does not  predict saturation of  the gluon distribution  in the very small $x$ limit.

A comprehensive phenomenological analysis of the behavior of distinct observables  for the HERA kinematical region using the Glauber-Mueller approach was made in Refs. \cite{ayala3,vic1,vic2}. In this kinematical region the solutions from the AGL equation and the Glauber-Mueller formula approximately coincide. 
The results from these analysis agree with the recent HERA data and allows to make some predictions which will be tested in a near future. Our main conclusion was that the unitarity corrections cannot be disregarded in the HERA kinematical region.

The unitarity corrections in the leading logarithmic limit can be studied using the dipole picture.
In this picture the unitarity corrections (multiple interactions between the onia) become important at high energies.
Recently,  an equation (K equation) which includes all pomeron exchanges in the leading logarithmic approximation using the dipole picture was proposed by Yu. Kovchegov \cite{kov}. In this contribution we present  the main steps of the demonstration that the K equation reproduces    the AGL equation and, consequently,  the GLR equation in the DLA limit.

\section{THE UNITARITY CORRECTIONS}

We start considering  the interaction between a virtual colorless hard probe and the nucleus via a gluon pair component of the virtual probe. To estimate the unitarity corrections we have to take into account the rescatterings of the gluon pair inside the nucleus. Using the Glauber-Mueller approach we consider the interations of the fastest partons with the target. As in QCD we expect that all partons from the cascade interact with the target, a generalized equation was proposed in   \cite{ayala1}. The AGL equation  is given by
\begin{eqnarray}
\partial^2_{y z} xG_A(x,Q^2)= 
 \frac{2\,Q^2}{\pi^2}  
\int \frac{d^2b_t}{\pi} [1 - e^{-\frac{1}{2}\sigma_N^{gg}(x
,Q^2)S(b_t)}] \nonumber
\label{agl}
\end{eqnarray}
where $y = ln(1/x)$, $z = ln(Q^2/\Lambda_{QCD}^2)$. This equation predicts the evolution of the nuclear gluon distribution $xG$ in the small $x$ region and is valid in the double logarithmic approximation (DLA).

As our goal  is to obtain the AGL from the dipole picture, we should make a transformation of the AGL equation  for $q\overline{q}$ dipoles, since in the dipole picture these dipoles are the basic configurations.
Moreover, considering a central collision ($b=0$) and that the  transverse cross-sectional area of the nucleus is $S_{\bot} = \pi  R^2$ and that $S(0) = A/(\pi R^2)$, the AGL equation for $b=0$ is obtained as
\begin{eqnarray}
\partial^2_{y z}xG_A(x,Q^2) = \frac{\cal{C}}{\pi^3}
   Q^2[1 - e^{-\frac{2\alpha_s \pi^2 }{N_c S_{\bot}} \frac{1}{Q^2
} xG_A(x,Q^2)}]\,, \nonumber 
\label{aglkov}
\end{eqnarray}
where ${\cal{C}} = N_c \,C_F \, S_{\bot}$. This equation takes into account that each parton in the parton cascade interacts with several nucleons within the nucleus (Glauber multiple scattering).

The GLR equation can be obtained directly from the AGL equation. If we expand the right hand side of this equation to the second order in $xG_A$ we obtain
\begin{eqnarray}
\partial^2_{y z} xG_A(x,Q^2)  = \frac{\alpha_s N_c}{\pi} \,xG_A(x,Q^2) \nonumber \\
- \frac{\alpha_s^2 \pi}{S_{\bot}} \frac{1}{Q^2} [xG_A(x,Q^2)]^2 \,\,, \nonumber
\label{glr}
\end{eqnarray}
which is the GLR equation for a cylindrical nucleus case. Moreover, if the unitarity corrections are small, only the first order in $xG_A$ contributes. In this limit the AGL equation  matches with the DGLAP evolution equation in the DLA limit.

Considering the multiple pomeron exchanges, Kovchegov has obtained an evolution equation for the total cross section  of the  $q\overline{q}$ pair with a transverse size $x_{01}$   interacting with the nucleus $N(\vec{x_{01}}, \vec{b_0}, Y)$ in the leading logarithmic approximation. The K equation was obtained considering the scattering of a virtual photon with a nucleus. The physical picture for this interaction is the same as the Glauber-Mueller approach. The incoming virtual photon generates a  $q\overline{q}$ pair which develops a cascade of gluons, which then scatters on the nucleus. In the large $N_c$ limit the gluon can be represented as a $q\overline{q}$ pair. Therefore, in this limit  and in the leading logarithmic approximation, the cascade of gluons can be interpreted as a dipole cascade, where each dipole in the cascade interacts with several nucleons within  the nucleus. Therefore, as the  K equation  and the AGL equation, although with distinct basic objects, resums the multiple scatterings of its degree of freedom, we expect that both coincide in a common limit.
In the double logarithmic limit the K equation reduces to
\begin{eqnarray}
\partial^2_{Y \epsilon} N(\vec{x_{01}}, \vec{b_0}, y) = 
\frac{\alpha_s C_F}{\pi}\,[2  -  N] N\,, \nonumber
\label{kovdif}
\end{eqnarray}
where $N = N(\vec{x_{01}}, \vec{b_0}, y)$, $y = ln(1/x)$ and $\epsilon = ln \,(1/x_{01}^2 \Lambda_{QCD}^2)$. We denote the above expression K (DLA) equation.

In \cite{kov} the connection between the quantity  $N(\vec{x_{01}}, \vec{b_0}, y)$ and the nuclear gluon distribution was discussed.  As there is some freedom in the definition of the gluon distribution, the choice for the connection between the two functions was arbitrary. Here we use the result obtained from  the nuclear structure function to establish the relation between these functions. The nuclear structure function in the Glauber-Mueller approach is given by 
\begin{eqnarray}
F_2^A(x,Q^2) =  \frac{Q^2}{4 \pi \alpha_{em}} R^2 \int dz \int \frac{d^2r_t}{\pi} |\Psi(z,r_t)|^2 \, \nonumber \\
\times   
\,2\,\{1 - e^{-\frac{ \alpha_s C_F \pi^2}{N_c^2 S_{\bot}}r_t^2 AxG(x,1/r_t^2)}\}\,\,. \nonumber
\label{f2eik3}
\end{eqnarray}
This expression estimates the unitarity corrections to the nuclear structure function for central collisions ($b=0$) in the DLA limit.

Consequently,  the total cross section of the $q\overline{q}$ pair interacting with the nucleus, $N(\vec{x_{01}}, \vec{b_0} = 0, Y)$, is given by
\begin{eqnarray}
N(\vec{x_{01}}, \vec{b_0} , y) = \,2\,\{1 - e^{-\frac{ \alpha_s C_F \pi^2}{N_c^2 S_{\bot}} x_{01}^2 AxG(x,1/x_{01}^2)}\}\,. \nonumber
\label{ene}
\end{eqnarray}
Substituting the  above expression in the K (DLA) equation and using $x_{01} \approx 2/Q$, the AGL equation is a straithforward consequence \cite{agl}.

\section{CONCLUSION}
We demonstrate in this contribution that the AGL equation can be obtained from the dipole picture.This result shows that the AGL equation is a good candidate for the unitarized evolution equation at small $x$ in the DLA limit, supported by two different frameworks describing small $x$ phenomena.

Another candidate for the unitarized evolution equation was proposed by Jalilian-Marian {\it et al.} \cite{jamal}. These authors have derived a general evolution equation for the gluon distribution  in the limit of large parton densities and leading logarithmic approximation, considering  a very large nucleus. This work is based on  effective Lagrangian formalism for the low $x$ DIS  and the Wilson renormalization group.
 In the general case the evolution equation  is a very complicated equation, which does not allow to obtain analytical solutions. Recently, these authors have considered the DLA limit on their result \cite{jamal1} and have shown 
that the evolution equation reduces to an equation with a functional form similar, but not identical, to the AGL equation. We believe that a more detailed analysis of the approximations used in both equations will allow to  demonstrate the equivalence of both equations in a common limit.

\end{document}